\documentclass[12pt]{article}
\usepackage{epsfig}
\usepackage{latexsym}
\newcommand{\be}{\begin{equation}}
\newcommand{\ee}{\end{equation}}
\newcommand{\bea}{\begin{eqnarray}}
\newcommand{\eea}{\end{eqnarray}}

\newcommand{\pl}{{\rm pl}}

\begin{document}
\baselineskip=16pt

\begin{center}
{\Large{\bf Is Our Universe Natural?}\footnote{Invited review
for {\it Nature}.  This work was supported in 
part by the U.S. Dept. of Energy, the National Science Foundation,
and the David and Lucile Packard Foundation.}}

\vspace*{0.1in}
Sean M.\ Carroll
\vspace*{0.1in}

\it Enrico Fermi Institute, Department of Physics, and\\
Kavli Institute for Cosmological Physics,
University of Chicago\\
5640 S.~Ellis Avenue, Chicago, IL~60637, USA\\
{\tt carroll@theory.uchicago.edu} \\
\vspace*{0.2in}
\end{center}

\baselineskip=17pt

\noindent {\bf It goes without saying that we are stuck with the
universe we have.  Nevertheless, we would like to go
beyond simply describing our observed universe, and try to 
understand why it is that way rather than some other way.
Physicists and cosmologists have been exploring increasingly
ambitious ideas that attempt to explain why certain features of
our universe aren't as surprising as they might first appear.}

What makes a situation ``natural''?  Ever since Newton,
we have divided the description of physical systems into two parts:
the configuration of the system, characterizing the particular state
it is in at some specific time, and the dynamical laws governing its
evolution.  For either part of this description, we have an intuitive
notion that certain possibilities are more robust than others.
For configurations, the concept of entropy quantifies how likely
a situation is.  If we find a collection of gas molecules 
in a high-entropy state distributed uniformly in a box, we are not
surprised, whereas if we find the molecules huddled in a
low-entropy configuration in one corner of the box we imagine there
must be some explanation.

For dynamical laws, the concept of naturalness can be harder to quantify.
As a rule of thumb, we
expect dimensionless parameters in a theory (including ratios of
dimensionful parameters such as mass scales) to be of order unity, not too
large nor too small.  Indeed, in the context
of effective quantum field theories, the renormalization group gives us
some justification for this notion of ``naturalness'' \cite{thooft}.  
In field theory, the dynamics of the low-energy degrees of freedom
fall into universality classes that do not depend on the detailed
structure of physics at arbitrarily high energies.  If an interaction
becomes stronger at large distances, we expect it to be relevant at
low energies, while interactions that become weaker are irrelevant;
anything else would be deemed unnatural.\footnote{The
parallel between natural states being 
high-entropy and natural theories arising from the renormalization group
is essentially an analogy, although there has been some tentative work
towards establishing a more formal connection --
in particular, linking Boltzmann's $H$-theorem
describing the evolution of entropy and the $c$-theorem describing
renormalization-group flow \cite{Zamolodchikov:1986gt,Cardy:1988cw}.
See for example \cite{Gaite:1995yg}.} 

If any system should be natural, it's the universe.  Nevertheless,
according to the criteria just described, the universe we observe
seems dramatically unnatural.  The entropy of the universe isn't nearly
as large as it could be, although it is at least increasing; for some
reason, the early universe was in a state of incredibly low entropy.
And our fundamental theories of physics involve huge hierarchies 
between the energy scales\footnote{Throughout this paper I will use
units in which $\hbar=c=1$, so that $1 {\rm ~eV} = 1.8\times 10^{-33}
{\rm ~g} = 5.1 \times 10^4 {\rm ~cm}^{-1} = 1.5 \times 10^{15}
{\rm sec}^{-1}$.} characteristic of gravitation (the reduced Planck
scale, $E_\pl = 1/\sqrt{8\pi G} \sim 10^{27}$ electron volts), particle 
physics (the Fermi scale of the weak interactions, $E_{\rm F} 
\sim 10^{11}$~eV, and the scale of quantum chromodynamics,
$E_{\rm QCD} \sim 10^{8}$~eV), and the recently-discovered vacuum energy 
($E_{\rm vac} \sim 10^{-3}$~eV).\footnote{We will not consider neutrino
masses, which have not yet been accurately measured.}
Of course, it may simply be that
the universe is what it is, and these are brute facts we have to live
with.  More optimistically, however, these apparently delicately-tuned
features of our universe may be clues that can help guide us
to a deeper understanding of the laws of nature.

I will survey some recent ideas for confronting these problems of
naturalness, both with respect to the state of the universe and the
laws of physics.  The choice of ideas to consider is completely
ideosyncratic, but should serve to give a flavor for the types of
scenarios being considered in the more speculative corners of
contemporary cosmology.

\begin{table}
\centering
\begin{tabular}{|l||c|c|}
\hline
Scale & Energy & Length \cr
\hline 
Planck (Gravitation) & ~~ $E_\pl = (8\pi G)^{-1/2}\sim 10^{27}$~eV~~ & 
~~ $L_\pl \sim 10^{-32}$~cm~~ \cr
Fermi (Weak interactions) & $E_{\rm F}= (G_{\rm F})^{-1/2} \sim 10^{11}$~eV 
& $L_{\rm F} \sim 10^{-16}$~cm \cr
QCD (Strong interactions) & $E_{\rm QCD} = \Lambda_{\rm QCD} \sim 10^{8}$~eV  
& $L_{\rm QCD} \sim 10^{-13}$~cm \cr
Vacuum (Cosmological constant) & $E_{\rm vac}= \rho_{\rm vac}^{1/4} 
\sim 10^{-3}$~eV  & $L_{\rm vac} \sim 10^{-2}$~cm  \cr
Hubble (Cosmology) & $E_{H} = H_0 \sim 10^{-33}$~eV  
& $L_{H} \sim 10^{28}$~cm \cr
\hline
\end{tabular}
\caption{Orders of magnitude of the characteristic scales of our universe,
in units where $\hbar = c = 1$.  The reduced Planck energy is derived
from Newton's constant $G$; the Fermi scale of the weak interactions
is derived from Fermi's constant $G_{\rm F}$.  The QCD scale 
$\Lambda_{\rm QCD}$ is the energy at which the strong coupling constant
becomes large.  The vacuum energy scale arises from the energy density
in the cosmological constant.  The Hubble scale characteristic of cosmology
is related to the total density $\rho$ by the Friedmann equation,
$H \sim \sqrt{\rho}/E_\pl$.  Note the large dynamic range spanned
by these parameters.}
\end{table}

\noindent{\bf The state of our universe.}

Consider the state in which we find our observable universe.  On
very large scales the distribution of matter is approximately 
homogeneous and isotropic, and distant galaxies are expanding away
from each other in accordance with Hubble's law.  Extrapolating into
the past, the universe originated in a hot, dense state about fourteen
billion years ago.  The deviations from perfect smoothness that we
observe today have grown via gravitational instability from 
initially small perturbations of approximately equal
amplitude on all length scales.  Interestingly, the ``ordinary matter''
made of particles described by the Standard Model of particle physics
is only about 4\% of the total energy of the universe;  another
23\% is some ``dark matter'' particle that has not yet been 
discovered.  Even more interestingly, 73\% of 
the universe is ``dark energy,'' some form of smoothly-distributed
and nearly-constant energy density
\cite{Riess:1998cb,Perlmutter:1998np,Spergel:2003cb}.
The most straightforward candidate for dark energy is vacuum energy,
or the cosmological constant:  an absolutely constant minimum energy
of empty space itself, with a density $\rho_{\rm vac} \sim
E_{\rm vac}^4 = (10^{-3} {\rm ~eV})^4$ \cite{Carroll:2000fy}.

If the universe were in a ``likely'' configuration, we would expect it
to be in a high-entropy state:  {\it i.e.}, in thermal equilibrium.
Unfortunately, we do not have a rigorous definition of entropy for
systems coupled to gravity, but we can make simple estimates.  The
entropy of matter and radiation (neglecting gravity) in our 
observable universe is approximately the number of massless particles,
$S_M(U) \sim 10^{88}$.  This was the dominant contribution at early times
when the matter distribution was essentially smooth; today, however, 
inhomogeneities have formed through gravitational collapse, creating
large black holes at the centers of galaxies, and
these black holes dominate the current entropy \cite{penrose}.  
The Bekenstein-Hawking entropy of a black hole is proportional to its
horizon area,  $S_{BH} = A/ 4G \sim 10^{77}(M_{BH}/M_\odot)^2$.
Since there are probably more than ten billion galaxies in the 
observable universe with million-solar-mass black holes at their
centers, the current entropy in black holes is at least
$S_{BH}(U) \sim 10^{99}$.  If we were to combine all of the matter
in the observable universe into one giant black hole, the entropy would 
be significantly larger, $S_{\rm max} \sim 10^{120}$.  So our universe 
is currently in a rather low-entropy configuration, and it started out 
much lower.\footnote{The fact that entropy tends to increase is of course 
just the Second Law of Thermodynamics, and makes sense once we assume that 
the initial entropy was low.  The puzzle is why the early universe should
differ so dramatically from the late universe, despite the intrinsic
time-reversibility of the microscopic laws of physics; this is known as the
``arrow of time'' problem.} 
 
Faced with the question of why our universe started out with such a
low entropy, most cosmologists would appeal to inflation
\cite{Guth:1980zm,Linde:1981mu,Albrecht:1982wi,Linde:1983gd}.  
According to this
idea, a very tiny patch of space dominated by false vacuum energy
undergoes a period of rapidly accelerated expansion, smoothing out any
inhomogeneities and ultimately reheating to the radiation-dominated
early state of the conventional Big Bang model.  For inflation to
begin, the initial patch must be quite smooth itself
\cite{Vachaspati:1998dy}; but because it is so small, one imagines that
it can't be that hard to find an appropriate region somewhere in the
chaotic conditions of the early universe.  

It is worth emphasizing that the {\it only} role of inflation
is to explain the initial conditions of the
observable universe.  And at this it does quite a good job;
inflation predicts that the universe should be spatially flat, 
and should have a scale-free spectrum of adiabatic density 
perturbations
\cite{Guth:1982ec,Hawking:1982cz,Starobinsky:1982ee,Bardeen:1983qw},
both of which have been verified to respectable precision by
observations of the cosmic microwave background \cite{Spergel:2003cb}.
But we are perfectly free to imagine that these features are simply
part of the initial conditions -- indeed, both spatial flatness
and scale-free perturbations were investigated long before inflation.
The only reason to invoke inflation is to provide a reason why such
an initial condition would be {\it natural}.

However, as Penrose and others have argued, there is a skeleton in
the inflationary closet, at least as far as 
entropy is concerned \cite{penrose,Hollands:2002yb,Albrecht:2004ke}.  
The fact that the initial proto-inflationary
patch must be smooth and dominated by dark energy implies that it must
have a very low entropy itself; reasonable estimates range from
$S_I \sim 10^{0} - 10^{20}$.  Thus, among randomly-chosen initial
conditions, the likelihood of finding an appropriate proto-inflationary
region is actually much {\it less} than simply finding the conditions
of the conventional Big Bang model (or, for that matter, of our
present universe).  It would seem that the conditions required 
to start inflation
are less natural than those of the conventional Big Bang.

\begin{figure}[t]
  \begin{center}
    \resizebox{8cm}{!}{\includegraphics{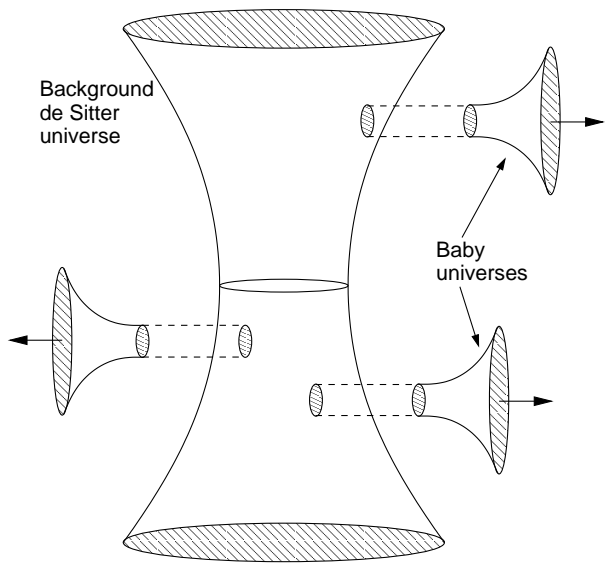}}
  \end{center}
  \caption{A possible spacetime diagram for the universe on ultra-large
   scales.  A natural state for a universe with a positive vacuum
   energy is empty de~Sitter space.  In the presence of an appropriate
   scalar field, quantum fluctuations in such a background can lead
   to the nucleation of baby universes.  Each baby universe is created
   in a proto-inflationary state, which then expands and reheats into
   a universe like that we observe.}
  \label{branefig}
\end{figure}

One possible escape from this conundrum has been suggested under the
name of ``spontaneous inflation'' \cite{Carroll:2004pn}, a particular
implementation of the idea of ``eternal inflation''
\cite{Vilenkin:1983xq,Linde:1986fc,Linde:1986fd,Goncharov:1987ir}
(for related ideas see \cite{Garriga:1997ef,Dutta:2005gt,Holman:2005ei}).
In general relativity, high-entropy states correspond to 
{\it empty space}; given any configuration of matter, we can always
increase the entropy by expanding the universe and diluting the matter
degrees of freedom \cite{Carroll:2004pn}.  But if empty
space has a nonzero vacuum energy, it can be unstable to the
creation of baby universes  \cite{Farhi:1986ty,Vilenkin:1987kf,
Farhi:1989yr,Fischler:1989se,Fischler:1990pk,Linde:1991sk}.
In an empty universe with a positive cosmological constant (de~Sitter
space), quantum fluctuations will occasionally drive scalar fields
to very large values of their potentials, setting up precisely the
conditions necessary to begin inflation.  The resulting bubble can
branch off into a disconnected spacetime, leading to an inflating
region that would resemble our observable universe. 

Needless to say, scenarios of this type are extremely speculative,
and may very well be completely wrong.  One mysterious aspect is the
process of baby-universe creation, which certainly lies beyond the
realm of established physics.  In the context of string theory, there
seem to be circumstances in which it can happen
\cite{Alberghi:1999kd,Adams:2005rb,Dijkgraaf:2005bp,McGreevy:2005ci}
but also ones in which it cannot \cite{Freivogel:2005qh}; whether it
will occur in realistic situations is still unclear.
Another issue is the nature of gravitational degrees of freedom
in the context of the holographic principle  
(see {\it e.g.} \cite{Dyson:2002pf}).
Given the relevance of inflation to observable phenomena, however,
it is important to establish that inflation actually solves the
problems it purports to address.  In spontaneous inflation, there is
a simple explanation for the low entropy of the initial state: it's
not really ``initial,'' but rather arises via quantum fluctuations
from a pre-existing de~Sitter
state with very large entropy and a very low
entropy density.  Whether this particular idea is on the right track
or not, it is crucial to understand whether inflation plays a role
in explaining how our observed configuration could be truly natural.

\noindent{\bf The laws of physics.}

The dynamical laws of nature at the microscopic level (including
general relativity and the Standard Model of particle physics) are
tightly constrained in the form that they may take, largely by
symmetry principles such as gauge invariance and Lorentz invariance.
The specific values of the numerical parameters of these theories
are in principle arbitrary, although on naturalness grounds we
would expect mass/energy scales to be roughly comparable to each
other.

As mentioned in the introduction, however, that is not what we
observe:  the Planck scale, Fermi scale, QCD scale, and vacuum-energy
scale are separated by huge hierarchies.  In the case of QCD this is
understandable; the characteristic scale is governed by the logarithmic
running of the QCD coupling constant, so a hierarchy isn't that
surprising.  The difference between the Planck and Fermi scales 
($E_\pl/E_{\rm F} \sim 10^{16}$) is
a celebrated puzzle in high-energy physics, known simply as ``the
hierarchy problem''; ideas such as supersymmetry may provide partial
answers.  The discrepancy between the Planck scale and the
vacuum-energy scale ($E_\pl/E_{\rm vac} \sim 10^{30}$) is another
puzzle, ``the cosmological constant problem''; most researchers agree
that there are no compelling solutions currently on the market.

In contemplating the nature of these hierarchies, a complicating factor
arises:  we couldn't exist without them.  If all of the energy scales
of fundamental physics were approximately equal, it would be impossible
to form structures in which quantum gravity didn't play a significant
role, and the large cosmological constant would make it difficult to
imagine any complex structures at all (since the rapidly-accelerating
expansion of the universe would work to tear things apart).  One can
imagine two very different attitudes in the face of this situation:
\begin{enumerate}
\item We just got lucky.  The constants of nature just happen to 
take on values consistent with the existence of complex organisms.
\item Environmental selection.
The parameters we observe are not truly fundamental, but merely
reflect our local conditions; hospitable regions of the universe will
always exhibit large hierarchies.
\end{enumerate}
Within the first case, there are two separate possibilities:  either we
are {\it really} lucky, in the sense that the observed hierarchies are
truly unnatural and have no deeper explanation, or there exist
unknown dynamical mechanisms which make these hierarchies perfectly
natural.  The latter possibility is obviously more attractive, although
it is hard to tell whether such dynamical explanation will eventually
be forthcoming.

Environmental selection, sometimes discussed in terms
of the ``anthropic principle,'' has received renewed attention since 
the discovery of the dark energy.  The basic idea is undeniably
true:  if our observable universe is only a tiny patch of a much
larger ``multiverse'' with a wide variety of local environments,
there is a selection effect due to the fact that life
can only arise in those regions that are hospitable to the existence
of life.  Of course, to give this tautology any explanatory 
relevance, it is necessary to imagine that such a multiverse actually
exists.

Existence of the multiverse requires two conditions:  the 
{\it possibility} of multiple vacuum states with differing
values of the constants of nature, and the {\it realization} of those
states in distinct macroscopic regions.  Recent ideas in string theory
suggest that there may be a ``landscape'' of metastable
vacua arising from different ways of compactifying extra dimensions
in the presence of branes and gauge fields -- numbers such
as $10^{500}$ vacua have been contemplated, promising more than 
enough diversity
of possible local conditions \cite{Bousso:2000xa,
Feng:2000if,Giddings:2001yu,Kachru:2003aw,Douglas:2003um,Ashok:2003gk}.
Meanwhile, the possibility of eternal inflation discussed in the
previous section provides a mechanism for realizing such states:
quantum fluctuations can lead to episodes of inflation that reheat
into universe-sized domains in any of the permitted vacuum states
\cite{Linde:2002gj}.
Thus, the ingredients of the multiverse scenario seem to be in place,
even if they remain quite speculative.

If all the multiverse does is {\it allow} for the existence of a
region that resembles our own, it adds nothing to our 
understanding; it is equally sensible
to say that our universe is simply like that.  Instead, the
possible epistemological role of the multiverse is to explain why
our observed parameters are natural.  In principle, the multiverse
picture allows us to {\it predict} the probability distribution
for these parameters.  In particular, the probability $P(X)$ that an
observer measures their universe to have feature $X$ (such as
``the ratio of the vacuum energy scale to the Planck scale is
of order $10^{-30}$'') will be roughly of the form
\be
  P(X) = \frac{\sum_{n} \sigma_n(X) V_n \rho_n}{\sum_{n}V_n \rho_n}\,.
  \label{px}
\ee
In this expression, the index $n$ labels all possible vacuum states;
$\sigma_n(X)$ equals $1$ if vacuum $n$ has property $X$ and $0$ if it
does not; $V_n$ is the spacetime volume in vacuum $n$; and $\rho_n$
is the spacetime density of observers in vacuum $n$.

Obviously, there is some imprecision in the definition of (\ref{px});
for example, we have been vague about what constitutes an 
``observer.''  (For attempts at more precise formulations of an 
equivalent expression, see \cite{Vilenkin:1994ua,
Tegmark:2004qd,Aguirre:2005cj,Garriga:2005av,Easther:2005wi,
Tegmark:2005dy}.)    This expression
suffices for our current purpose, however, which is to point out
that actually calculating the probability is at best beyond our
current abilities, and at worst completely hopeless.  Just to mention the
most obvious difficulty:  in the context of eternal inflation, there is
every reason to believe that the volumes $V_n$ of some (if not all)
vacua are infinite, and the expression is simply undefined.

The fact that (\ref{px}) is undefined hasn't stopped people from trying
to calculate it.  The most famous example is Weinberg's prediction of
the magnitude of the cosmological constant
\cite{Weinberg:1987dv,Martel:1997vi,Banks:2000pj}.
This calculation imagines a flat prior distribution for the vacuum
energy, keeping all other parameters fixed, and relies on the fact that
the universe recollapses in the presence of a large negative vacuum
energy and expands too quickly for galaxies to form in the presence of
a large positive vacuum energy.  Under these assumptions, the predicted
value of $\rho_{\rm vac}$ is not too different from what has been
observed; moreover, the prediction was made before the measurement.
Attempts have also been made to apply similar reasoning to 
models of particle physics
\cite{Pogosian:2004hd,Arkani-Hamed:2004fb,Arkani-Hamed:2005yv,
Dine:2005yq,Fox:2005yp}.

Unfortunately, there is little reason to be satisfied with this
calculation of the expected vacuum energy.
In terms of (\ref{px}), it is equivalent to imagining that the
factor $\sigma_n(X)$ counting appropriate vacua is distributed uniformly
in $\rho_{\rm vac}$, the volume term $V_n$ is simply a 
constant, and the density of observers $\rho_n$ is proportional to the 
number of galaxies.  The first of these is a guess, the second is
likely to be fantastically wrong in the context of eternal inflation,
and the last only makes sense if all of the other parameters are held
fixed, which is not how we expect the multiverse to work.  For example,
allowing the amplitude of primordial density fluctuations to vary along
with the vacuum energy can dramatically change the result
\cite{Tegmark:1997in,Graesser:2004ng,Garriga:2005ee,Feldstein:2005bm}.

At the present time, then, there is {\it not} a reliable 
environmental explanation
for the observed value of the cosmological constant.  
Meanwhile, other attempts to use anthropic reasoning
lead to predictions that are in wild disagreement with observations
\cite{Olum:2003ma}.  But objections to the credibility of the
apparently-successful predictions of the multiverse idea have equal
force when applied to the apparently-unsuccesful predictions; whether
or not the idea is falsifiable, it would be an exaggeration to say
that it's already been falsified.  

More importantly, limitations in our current ability to calculate
expectation values in the multiverse are not evidence that there isn't
some truth to the idea itself.  If we eventually decide that environmental
selection plays no important role in explaining the observed parameters
of nature, it will be because we somehow come to believe that the 
parameters we measure locally are also characteristic of regions beyond 
our horizon, not because the very concept of the multiverse is
aesthetically unacceptable or somehow a betrayal of the
Enlightenment project of understanding nature through reason and
evidence.

\noindent{\bf Discussion.}

Naturalness is an ambiguous guide in the quest to better understand
our universe.  The observation that a situation seems unnatural within
a certain theoretical context does not carry anything like the 
force of an actual contradiction between theory and 
experiment.  And despite our best efforts, naturalness is something
that is hard to objectively quantify.

The search for naturalness plays an important role as a {\it hint} of
physical processes that are not yet understood.  In particle physics,
attempts to find a natural solution to the hierarchy problem have
driven investigations into supersymmetry and other models; in cosmology,
attempts to explain the uniformity and flatness of our contemporary
universe helped drive the development of the inflationary scenario.
We can hope that attempts to understand the cosmological constant and
the low entropy of the early universe will lead to compelling new ideas 
about the fundamental architecture of nature.

The scenarios discussed in this paper involve the invocation of multiple
inaccessible domains within an ultra-large-scale multiverse.  For good
reason, the reliance on the properties of unobservable regions and the
difficulty in falsifying such ideas makes scientists reluctant to grant 
them an explanatory role \cite{Smolin:2004yv}.  Of course, the idea
that the properties of our observable domain can be uniquely extended
beyond the cosmological horizon is an equally untestable assumption.
The multiverse is not a theory; it is
a {\it consequence} of certain theories (of quantum gravity and cosmology),
and the hope is that these theories eventually prove to be testable in
other ways.  Every theory makes untestable predictions, but theories
should be judged on the basis of the testable ones.  The ultimate
goal is undoubtedly ambitious:  to 
construct a theory that has definite consequences for
the structure of the multiverse, such that this structure provides an
explanation for how the observed features of our local domain can arise
naturally, and that the same theory makes predictions that can be directly
tested through laboratory experiments and astrophysical observations.
Only further investigation will allow us to tell whether such a
program represents laudable aspiration or misguided hubris.


\end{document}